\documentclass[a4paper,11pt]{article}
\pdfoutput=1 

\usepackage{jheppub} 

\usepackage[T1]{fontenc} 

\usepackage{booktabs}
\usepackage{hyperref}
\usepackage{multirow} 
\usepackage{numprint}
\usepackage{enumitem}
\usepackage{mathrsfs}
\usepackage{graphicx}
\usepackage{multirow}
\usepackage{makecell}
\usepackage{verbatim}
\usepackage{color}
\usepackage{comment}
\usepackage{epstopdf}
\usepackage{float}
\usepackage{amsmath}
\usepackage{amssymb}
\usepackage{tikz}
\usepackage{wrapfig}
\usepackage{tikz-feynman}
\usepackage{verbatim}
\pdfoutput=1
\usepackage[utf8]{inputenc}
\usepackage{graphicx,amsmath,amsfonts,amssymb,jheppub,slashed,float}
\usepackage{bbold}
\usepackage{numprint}
\usepackage{enumitem}
\usepackage{graphicx,color,amsmath}
\usepackage{hyperref}
\usepackage{hyperref}
\usepackage{url}
\usepackage{slashed,stmaryrd}

\newcommand*{\dif}{\mathop{}\!\mathrm{d}}
\allowdisplaybreaks
\makeatletter
\newcommand{\xRightarrow}[2][]{\ext@arrow 0359\Rightarrowfill@{#1}{#2}}
\makeatother
\usepackage{pgfplots}

\usepackage{bm}
\usepackage{mathtools}

\renewcommand{\vec}[1]{\bm{#1}}
\usetikzlibrary{positioning,arrows,patterns}
\usetikzlibrary{decorations.markings}
\usetikzlibrary{calc}

	\title{Migdal effect of Phonon-mediated neutrino nucleus scattering in semiconductor detectors}
	\author[a,b]{Yu-Feng Li,\note{liyufeng@ihep.ac.cn}}
	\author[a,b]{Shuo-yu Xia,\note{xiashuoyu@ihep.ac.cn~(corresponding author)}}
	\affiliation[a]{Institute of High Energy Physics, Chinese Academy of Sciences, Beijing 100049, China}
	\affiliation[b]{School of Physical Sciences, University of Chinese Academy of Sciences, Beijing 100049, China}
 
\abstract{

In this work, we introduce the theoretical framework of the phonon-mediated Migdal effect for neutrino-nucleus scattering in semiconductors, considering both the Standard Model and the presence of the neutrino magnetic moment. We calculate the rate of electron excitation resulting from the Migdal effect and observe a substantial coherent enhancement compared to ordinary neutrino-electron scattering. Furthermore, we provide numerical sensitivities for detecting the Migdal effect and constraints on the neutrino magnetic moment, utilizing an experimental setup with reactor neutrinos.

}

\begin{document}

\maketitle
\flushbottom                                                            
{
 
}
 
\section{Introduction}
The search for dark matter has seen tremendous progress in the past decade~\cite{SuperCDMS:2018mne,EDELWEISS:2020fxc,SENSEI:2020dpa,LZ:2022ufs,XENON:2018voc,XENON:2020fgj} and the direct detection experiments play crucial roles in searching for the dark matter elastic scattering on targets. There have been notable advances in the direct detection techniques with various target materials such as semiconductors~\cite{Liang:2021zkg,Liang:2018bdb,Graham:2012su,Essig:2015cda,Hochberg:2016sqx}, superfluid helium~\cite{Caputo:2019xum,Caputo:2019cyg,Knapen:2016cue} and superconductors~\cite{Hochberg:2015pha,Hochberg:2016ajh}. The direct detection experiments have shown a high degree of sensitivity to the nuclear recoil signals lower than 100 keV with rather high target masses and excellent shielding from cosmic ray muons. With these excellent features, the direct detection experiments, which share many similarities with underground neutrino detectors, can serve as useful tools for the precise test of the electroweak theory~\cite{Papoulias:2017qdn,Cadeddu:2018izq}, measuring the neutrino electromagnetic properties~\cite{Cadeddu:2018dux,Cadeddu:2019eta,Kim:2021lun} and searching for the evidence of nonstandard neutrino interactions beyond the Standard Model (SM)~\cite{Coloma:2017ncl, Liao:2017uzy,Denton:2018xmq, AristizabalSierra:2018eqm, Giunti:2019xpr}. Furthermore, direct detection experiments with extremely low energy thresholds like SENSEI~\cite{SENSEI:2020dpa} and EDELWEISS~\cite{EDELWEISS:2017uga} may serve as excellent platforms to explore novel neutrino interactions at the very low energy scale.

The Migdal effect was first proposed by A.B. Migdal in 1939~\cite{migdal:1939svj} as the process where the emission of ionized electrons from suddenly struck nucleus. Just recently in 2017, the authors of Ref.~\cite{Ibe:2017yqa} have provided a theoretical clarification that the Migdal effect can be triggered more easily than expected. Subsequently, the theoretical description of the Migdal effect has been significantly improved~\cite{Essig:2019xkx,Bell:2019egg,Dolan:2017xbu,Berghaus:2022pbu}.
In addition, the Migdal effect has garnered considerable attention in multiple physical fields including dark matter detection, nuclear physics and particle physics in recent years~\cite{Gelmini:2020xir,Knapen:2020aky,Liang:2020ryg,Liang:2019nnx,Liang:2022xbu,Essig:2019xkx}. Although a direct observation of the Migdal effect in the nucleus collision has not been achieved, various attempts and proposals to conduct the measurements are already under way, utilizing either the neutron nucleus scattering~\cite{Berghaus:2022pbu, Bell:2021ihi,Araujo:2022wjh, Cox:2022ekg,Adams:2022zvg,Xu:2023wev} or the dark matter direct detection experiments~\cite{EDELWEISS:2022ktt,LUX:2018akb, CDEX:2019hzn,EDELWEISS:2019vjv,XENON:2019zpr,SENSEI:2020dpa,Essig:2022dfa}. 
Neutrinos can be a potential tool to explore the parameter space of the Migdal effect, because the theory of standard neutrino-nucleus scattering is well established~\cite{Drukier:1984vhf,Patton:2012jr}, and meanwhile the electron excitation from the Migdal effect allows for distinguishing the corresponding events from coherent elastic neutrino-nucleus scattering (CE$\nu$NS) without the Migdal effect~\cite{Tiffenberg:2017aac}, which is a major component of the neutrino scattering signal at such low energies. However, the corresponding Migdal excitation still benefits from the coherent enhancement effect of the CE$\nu$NS process.

Semiconductor materials have significantly lower ionization thresholds of $\mathcal{O}(1)\;{\rm eV}$ compared to the typical ionization threshold in atoms of the order of $\mathcal{O}(10)\;{\rm eV}$. Therefore, semiconductors are considered as ideal targets for exploring the Migdal effect. Extensive efforts have been made to improve the theoretical understanding of the Migdal effect in semiconductors at low energy~\cite{Liang:2020ryg,Liang:2019nnx,Liang:2022xbu,Essig:2019xkx}. In particular, the authors of Ref.~\cite{Liang:2022xbu} have developed a multi-phonon model to describe the Migdal effect induced by non-relativistic dark matter particles in semiconductors at low energies, where the contribution of phonons cannot be neglected. However, a proper theoretical description of the Migdal effect induced by relativistic neutrinos in semiconductors, which is also an important observation goal in the dark matter direct detection experiments, is still lacking.
In this work, we will extend the framework of phonon-mediated Migdal effect in Ref.~\cite{Liang:2022xbu} to neutrino-nucleus scattering in semiconductor and calculate the corresponding event rate in silicon with reactor neutrino flux. The contribution of the Migdal effect with reactor neutrinos are investigated and we present the corresponding sensitivity with silicon as target. 
{The neutrino magnetic moment is estimated to be vanishingly small in simple extensions of the SM~\cite{PhysRevLett.45.963,Giunti:2014ixa}, while in many specific models beyond the simplest SM extension~\cite{Bell:2006wi,Bell:2005kz} it can be significantly enhanced and testable in current or future laboratory measurements~\cite{Amsler:2002tu,TEXONO:2002pra,Beda:2007hf,Giunti:2014ixa} and astrophysical considerations~\cite{PhysRevLett.111.231301,C_rsico_2014,Carenza:2022ngg,Li:2022dkc}.
Since the contribution of the neutrino magnetic moment in CE$\nu$NS is enhanced by the inverse of the recoil energy, those experiments with extremely low threshold are required to efficiently probe its nature. Therefore, semiconductor detectors take significant advantages compared to other popular detectors in this field and meanwhile we will explore the contribution of the Migdal effect in the presence of the neutrino magnetic moment in the present work.}


This paper is organized as follows. Section II provides a complete description of the theoretical framework of the neutrino-induced phonon-mediated Migdal effect and the corresponding Migdal excitation rate in silicon with reactor neutrinos as a source. In Section III, we analyze the associated sensitivity of the Migdal excitation as well as the detection potential of the neutrino magnetic moment with the Migdal effect based on the silicon target. Finally we give the concluding remarks and discussions in Section IV.

\section{Neutrino induced Migdal effect as a multi-phonon process}

In this section we derive the formulation of the cross section contributed by the Migdal effect induced by the CE$\nu$NS in semiconductor materials under the multi-phonon framework. Since the energy region we are concerned is a few tens of eV and the collective behavior of electrons and nuclei can not be neglected, the concept of a freely-recoiling nucleus no longer holds. However, the corresponding description in the context of phonons can be naturally extended to such low energies. 
To describe the neutrino induced Migdal effect at this energy region, the effects of the incident neutrino can be approximately regarded as a perturbation on the lattice system of the target material~\cite{Liang:2021zkg}. This approach has already been theoretically illustrated in Ref.~\cite{Li:2023jhep}, which utilizes the framework of the non-relativistic effective field theory (NR EFT) to describe the elastic neutrino-electron scattering (E$\nu$ES). The Lagrangian for CE$\nu$NS induced by reactor electron neutrinos can be written as
\begin{equation}
	\mathcal{L}_{\rm CE\nu NS}=-i\frac{\sqrt{2}}{2}G_{\rm F}[g_{n}NF_N^2(\vec{q})+g_{p}ZF_Z^2(\vec{q})]{\nu}_{e}\gamma^{\mu}(1-\gamma^{5})\bar{\nu}_{e}\bar{\mathcal{N}}\gamma_{\mu}\mathcal{N}\,,
\end{equation}
where $\mathcal{N}$ is the nucleus and $\bar{\nu}_{e}$ is the electron neutrino field from the nuclear fission of reactors. The constants to describe the couplings between the neutrino and neutron and proton are $g_{n}=1/2$ and $g_{p}=-1/2+2\sin^2 \theta_W$ respectively. Since the energy region we are concerned is lower than 1 keV, the coherence effects plays an important role in enhancing the interaction cross section and the neutron number $N$ and proton number $Z$ coherently contribute to the scattering process. The form factor $F_{N,Z}^2(\vec{q})$ can be also approximately considered as 1 at this energy region. With a similar way with that described in Ref~\cite{Li:2023jhep}, the above Lagrangian can be written as an effective one in the energy region that we are concerned
\begin{equation}
	\mathcal{L}_{\rm eff}=
	-i\frac{\sqrt{2}}{2}G_{\rm F}[N-Z(1-4\sin^2 \theta_W)]\nu_{e,L}\gamma^{0}\bar{\nu}_{e,L}\phi_{+}^{\dagger}\phi_{+}\,,
\end{equation}
where $\phi_{+}$ is the non-relativistic effective operator of the target nucleus. It can be proved that the above effective Lagrangian is identical for all flavor neutrinos. From this expression, we can calculate the corresponding effective potential that contributes to the perturbation on the nucleus system
\begin{equation}
	\mathcal{V}_{\rm CE\nu NS}=
	-i\frac{\sqrt{2}}{2}G_{\rm F}[N-Z(1-4\sin^2 \theta_W)]\,.
\end{equation}
Following the Feynman rules for the multi-phonon process summarized in Ref~\cite{Liang:2021zkg}, the amplitude of the neutrino induced Migdal effect as a multi-phonon process in the framework of the zero-temperature QFT can be written as
\begin{equation}
	\begin{aligned}
		i\mathcal{M}& =(-i)\mathcal{V}_{\rm CE\nu NS}(\bm{q})Ne^{-W(\bm{q})}\prod_{s=1}^{n}\left(\frac{-i\bm{q}\cdot\bm{\epsilon}_{\bm{k}_s,\alpha_s}}{\sqrt{2Nm_N\omega_{\bm{k}_s,\alpha_s}}} \right) \\
		&\times\sum_{\bm{G}^\prime}\sum_{\bm{k},\alpha}\left(\frac{-i\bm{q}\cdot\bm{\epsilon}_{\bm{k},\alpha}}{\sqrt{Nm_N}} \right) \left\lbrace \left[\frac{i}{(\varepsilon_i-\varepsilon_j)^2-\omega^2_{\bm{k},\alpha}} \right] \left[\frac{i(\bm{k}+\bm{G}^{\prime})\cdot\bm{\epsilon}_{\bm{k},\alpha}}{\sqrt{Nm_N}} \right]\right. \\ 
		&\left.\times\left[ \frac{-iNZ_{ion}4\pi\alpha_e}{V|\bm{k}+\bm{G}^\prime|^2}\right] <i|e^{i(\bm{k}+\bm{G}^\prime)\cdot\hat{\bm{x}}}|j>\right\rbrace
		\sum_{\bm{G}}\delta_{\sum_{s}\bm{k}_s+\bm{G}+\bm{k},\bm{G}}\, ,
	\end{aligned}
\end{equation}
where $m_N$ is the mass of the target nucleus and $e^{-W(\bm{q})}$ is the Debye-Waller factor~\cite{Debye:1913adp,Waller:1923zfp} that describes the effect of the quantum and thermal uncertainties of the lattice. The first line accounts for the process where the perturbation of the incident neutrino excites a multi-phonon excitation on the nucleus system with $\bm{q}$ is the transferred momentum. The remaining part accounts for the process where the phonon induced by the incident neutrino leads to the excitation of a band electron and explains the electron excitation in the Migdal effect in the view of a multi-phonon model. The momentum of electron is written in the form of a momentum $\bm{k}$ within a Bloch Zone (BZ) and reciprocal lattice vectors $\bm{G}$ and $\bm{G}^{\prime}$. Since the target material we have considered is the semiconductor, the band gap $\varepsilon_i-\varepsilon_j$ is much larger than the phonon eigenenergies $\omega_{\bm{k},\alpha}$ and the contraction relation $\sum_{\alpha}(\bm{q}\cdot\bm{\epsilon}_{\bm{k},\alpha})\cdot[(\bm{k}+\bm{G}^{\prime})\cdot\bm{\epsilon}_{\bm{k},\alpha}]=\bm{q}\cdot(\bm{k}+\bm{G}^{\prime})$ can be applied to simplify the amplitude
\begin{equation}
	\begin{aligned}
		i\mathcal{M}=&(-i\mathcal{V}_{\rm CE\nu NS}(\bm{q})Ne^{-W(\bm{q})}\prod_{s=1}^{n}\left(\frac{-i\bm{q}\cdot\bm{\epsilon}_{\bm{k}_s,\alpha_s}}{\sqrt{2Nm_N\omega_{\bm{k}_s,\alpha_s}}} \right) \\
		\times&\sum_{\bm{G}^\prime}\sum_{\bm{k}}\left(\frac{\bm{q}\cdot(\bm{k}+\bm{G}^{\prime})}{m_N} \right) \left[\frac{1}{(\varepsilon_i-\varepsilon_j)^2} \right] \left[ \frac{Z_{ion}4\pi\alpha_e}{V|\bm{k}+\bm{G}^\prime|^2}\right] <i|e^{i(\bm{k}+\bm{G}^\prime)\cdot\hat{\bm{x}}}|j>\\
		\times&\sum_{\bm{G}}\delta_{\sum_{s}\bm{k}_s+\bm{G}+\bm{k},\bm{G}}\, .
	\end{aligned}
\end{equation}
With the relations of the neutrino flux $\Phi(E_{\nu})$ and recoil energy $E_R(\bm{q})=q^2/(2m_N)$, the resulting event rate for the neutrino induced Migdal effect can be written as
\begin{equation}
	\begin{aligned}
		R&=\frac{N}{V^2}Z^2\sum_{\bm{q}}2|{\mathcal{V}_{\rm CE\nu NS}}|^2\int \frac{\dif \omega}{\omega^4}\frac{\dif\Phi(E_{\nu})}{\dif E_{\nu}}\frac{\dif E_{\nu} \dif \cos \theta}{2}\\
		&\times\sum_{\left\lbrace n_i\right\rbrace} \frac{e^{-\frac{E_R(\bm{q})}{3N\omega_{1}}}}{n_1!} \left( \frac{E_R(\bm{q})}{3N\omega_1}\right) ^{n_1}\dots\dots\frac{e^{-\frac{E_R(\bm{q})}{3N\omega_{{3N}}}}}{n_{3N}!}\left( \frac{E_R(\bm{q})}{3N\omega_{3N}}\right) ^{n_{3N}}\delta(\sum_{i=1}^{3N}n_i\omega_1+\omega_{pp^\prime})\\
		&\times \sum_{\bm{G}^{\prime},\bm{G}^{\prime\prime}}\sum_{\bm{k}^{\prime}\in1\rm{BZ}} \frac{\bm{q}\cdot(\bm{k}+\bm{G}^{\prime})}{m_N}\frac{\bm{q}\cdot(\bm{k}+\bm{G}^{\prime\prime})}{m_N}\rm{Im}\left[ \epsilon^{-1}_{\bm{G}^{\prime},\bm{G}^{\prime\prime}}(\bm{k},\omega)\right] \tilde{V}_{\bm{G}^{\prime},\bm{G}^{\prime\prime}}^{cou}(\bm{k})\, ,
	\end{aligned}
\end{equation}
where $n_i$ describes the occupation number of the energy $\omega_i$ and $\rm{Im}[ \epsilon^{-1}_{\bm{G}^{\prime},\bm{G}^{\prime\prime}}(\bm{k},\omega)]$ is the energy loss function of the material. The number of the valence electrons $Z$ depends on the material and $N$ is the number of nuclei in the target material. The effective Coulomb potential is defined as $\tilde{V}_{\bm{G}^{\prime},\bm{G}^{\prime\prime}}^{cou}(\bm{k})={4\pi\alpha}/{(|\bm{k}+\bm{G}^{\prime}|\cdot|\bm{k}+\bm{G}^{\prime\prime}|)}$ and $\omega$ is the deposited energy in the lattice. The second line is the scattering function $S(q,E)$, which can be rewritten in the following form
\begin{equation}
	\begin{aligned}
		S(q,E)=e^{-E_R(\bm{q})\sum_{i}\frac{1}{N\omega_{i}}} \sum_{+\infty}^{n=0}\frac{E_R(\bm{q})^n}{n!}T_n(-E_{\nu}^{\prime}+E_{\nu}-\omega)\, ,
	\end{aligned}
\end{equation}
where 
\begin{equation}
	\begin{aligned}
	T_n(\omega)&=\frac{1}{2\pi}\int_{-\infty}^{+\infty}f(t)^{n}e^{-i\omega t}\dif t\, ,\\
	f(t)&=\sum_{i=1}^{3N}\frac{1}{3N}\frac{e^{i \omega_it}}{\omega_i}\,.	\\
	\end{aligned}
\end{equation}
The scattering function $S(q,E)$ is a combined Poisson distribution and it can be proved that it converges to a Gaussian form at large $q$ region with the center at $E_R(q)$~\cite{Liang:2022xbu}.
To further simplify the calculation, we define $E=-E_{\nu}^{\prime}+E_{\nu}-\omega$ and with $E_{\nu}^{\prime}=\sqrt{E_{\nu}^2-2qE_{\nu}\cos \theta+q^2}$ the integration of $T_n(-E_{\nu}^{\prime}+E_{\nu}-\omega)$ can be rewritten as
\begin{equation}
	\begin{aligned}
		&\int \dif \cos \theta T_n(-\sqrt{E_{\nu}^2-2qE_{\nu}\cos \theta+q^2}+E_{\nu}-\omega)\\
		=&\int_{0}^{2E_\nu-q-\omega(q>E_\nu),\,q-\omega(q<E_\nu)} \dif E \left| \frac{\dif \cos \theta}{\dif E}\right| T_n(E)\\
		=&\int_{0}^{2E_\nu-q-\omega(q>E_\nu),\,q-\omega(q<E_\nu)} \dif E \frac{-E-\omega+E_\nu}{E_\nu q} T_n(E)\, ,
	\end{aligned}
\end{equation}
It can be found that $E_R(q)=q^2/2M$ is much smaller than the upper limit of the above integration under most conditions and the Gaussian function degenerates into a delta function $\delta(E-q^2/2M)$ as a result. Therefore, the event rate can be simplified as
\begin{equation}
	\begin{aligned}
		R&=\frac{2\alpha NZ^2}{3\pi m_N^2\Omega}|{\mathcal{V}_{\rm CE\nu NS}}|^2 \int {q^3\dif q}\int \frac{\dif \omega}{\omega^4}\mathcal{F}(\omega)\frac{\dif\Phi(E_{\nu})}{\dif E_{\nu}}{\dif E_{\nu}}\\
		&\times \int_{0}^{2E_\nu-q-\omega(q>E_\nu),\,q-\omega(q<E_\nu)}  \frac{-E-\omega+E_\nu}{E_\nu q}\delta(E-q^2/2M)\dif E \, ,
	\end{aligned}
\end{equation}  
where we have employed the the assumption that the target material is isotropic and $\Omega$ is the volume of the unit cell.
Following Ref.~\cite{Liang:2020ryg}, it is convenient to define the non-dimensional averaged energy loss function $\mathcal{F}(\omega)$, which takes the form of 
\begin{equation}
	\mathcal{F}(\omega)=\sum_{\bm{G}^{\prime},\bm{G}^{\prime\prime}}\int_{1\rm{BZ}}\frac{\Omega\dif^3 {k}}{(2\pi)^3} \frac{(\bm{k}+\bm{G}^{\prime})\cdot(\bm{k}+\bm{G}^{\prime\prime})}{|\bm{k}+\bm{G}^{\prime}|\cdot|\bm{k}+\bm{G}^{\prime\prime}|}\rm{Im}\left[ \epsilon^{-1}_{\bm{G}^{\prime},\bm{G}^{\prime\prime}}(\bm{k},\omega)\right] \, .
\end{equation}
Since the energies of the present neutrino source range from a few hundreds of keV to tens of MeV, it is reasonable to make the assumption that $E_\nu \gg q^2/2M$ and $E_\nu\gg \omega$ and then the event rate can be written as 
\begin{equation}
	\begin{aligned}
		R&=\frac{\alpha NZ^2}{3\pi m_N^2\Omega}G_F^2[N-Z(1-4\sin^2 \theta_W)]^2\int\frac{\dif\Phi(E_{\nu})}{\dif E_{\nu}}\dif E_{\nu} \int q^3\dif q\int \frac{\dif \omega}{\omega^4}\mathcal{F}(\omega)\, .
	\end{aligned}
	\label{Eq:Rate_SM}
\end{equation}
Specifically, in this work we will apply to the study of the neutrino magnetic moment, which has attracted considerable efforts over the past few decades. The effective potential contributed by the neutrino magnetic moment in CE$\nu$NS can be written in a similar way with that in E$\nu$ES, which can be derived as~\cite{Li:2023jhep}
{\color{black}
\begin{equation}
	\begin{aligned}
		\mathcal{V}_{\rm mag}=Z\frac{\mu_{\nu_{e}}}{\mu_B}\frac{E_{\nu}}{m_e}\frac{8\pi\alpha}{q^2}\, ,
	\end{aligned}
\end{equation} }
where $\mu_{\nu_{e}}$ is the magnetic moment of the electron neutrino {\color{black}and in the unit of $\mu_B$}. Since in the model we have employed the scattering process is regarded as a perturbation on the lattice system, we can calculate the event rate from the contribution of the neutrino magnetic moment by simply replacing the potential of the standard neutrino-nucleus scatting $\mathcal{V}_{\rm CE\nu NS}$ with $\mathcal{V}_{\rm mag}$ in the above derivation as
\begin{equation}
	\begin{aligned}
		R&=\mu_{\nu_{e}}^2\frac{128\pi N\alpha^3 Z^4}{3 m_N^2m_e^2\Omega}\int E_{\nu}^2\frac{\dif\Phi(E_{\nu})}{\dif E_{\nu}}\dif E_{\nu} \int \frac{\dif q}{q}\int \frac{\dif \omega}{\omega^4}\mathcal{F}(\omega)\, .
	\end{aligned}
	\label{Eq:Rate_mag}
\end{equation}
\begin{figure*}
	\centering
	\includegraphics[width=\textwidth, angle=0]{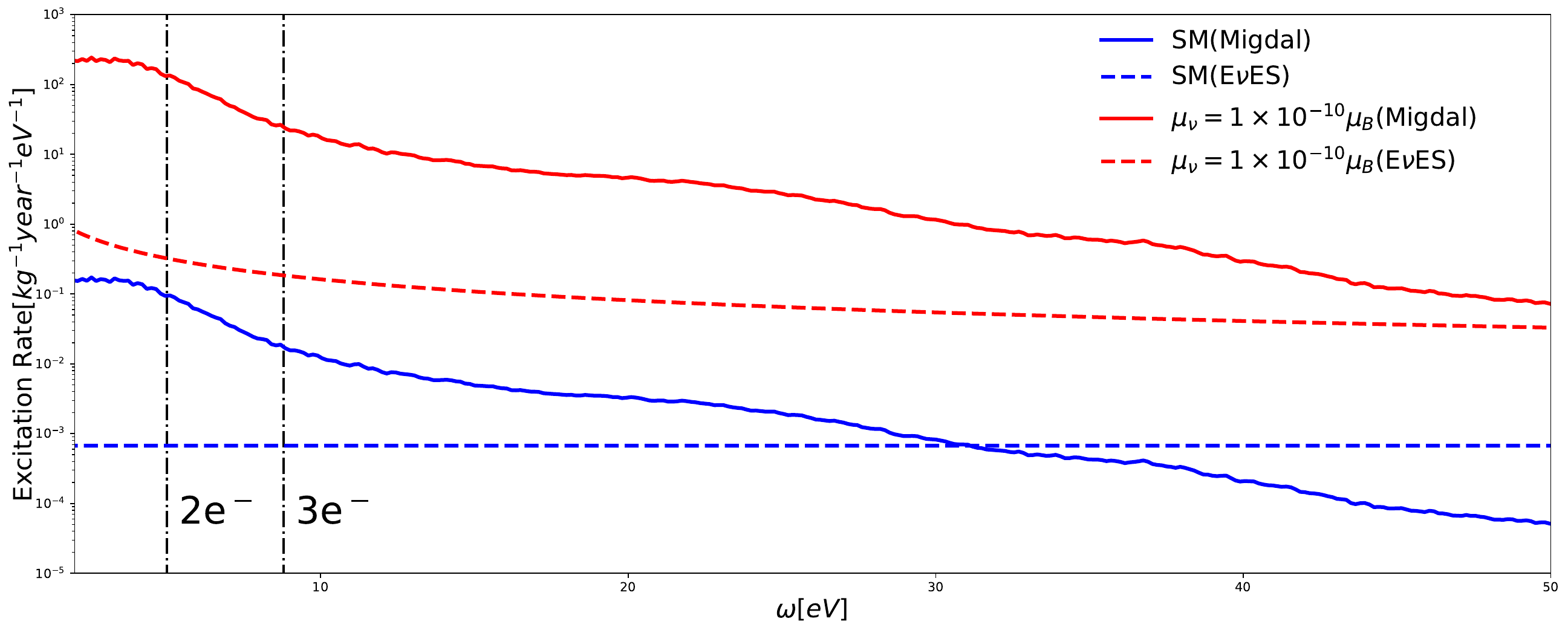}
	\caption{The differential excitation rates in the silicon target induced by reactor neutrinos with the neutrino magnetic moment $\mu_{\nu}=10^{-10}\mu_{\rm{B}}$ (red) and in the SM (blue). The neutrino source of Taishan-1 reactor and a distance of 30 meters are employed for the numerical calculations~\cite{JUNO:2020ijm}. The total excitation rates with the Migdal effect are shown with the solid lines and those from E$\nu$ES in dash lines. {\color{black}The mean energies for two and three ionization electron signals are shown with vertical dash dotted lines.}}
	\label{fig:spectra}	
\end{figure*}
For illustration, we display the differential excitation rates induced by reactor neutrinos via the E$\nu$ES and multi-phonon Migdal processes in  Fig.~\ref{fig:spectra}. We present the results for the silicon target under the SM (blue) and in the presence of the neutrino magnetic moment (red), which highlight the great potential of the multi-phonon Migdal process to explore new physics beyond the SM. The contributions from the Migdal effect are shown in solid lines and those from E$\nu$ES are shown in dash lines. For illustration, we choose the Taishan-1 reactor as the neutrino source and employ the distance at 30 meters, which is a similar set with that of Taishan Antineutrino Observatory (TAO)~\cite{JUNO:2020ijm} with the inverse-beta-decay interaction as the primary target. It should be noticed that the Migdal process exhibits a rapid decrease in the excitation rate with the increasing of the excited energy, limiting the capability of detectors to probe the relevant physics unless they can achieve an exceedingly low threshold. We also label the maximal mean energies for one and two ionization electron signals respectively, by using the mean energy per electron-hole pair $\epsilon=3.8$ eV, the band gap energy for silicon $E_{\rm gap}=1.2$ eV from Ref.~\cite{Berghaus:2022pbu}, and the relation 
\begin{equation}
	Q=1+\large\lfloor\frac{\omega-{E}_{\rm gap}}{\epsilon}\large\rfloor\,,
\end{equation}
where $Q$ is the number of ionization electrons {\color{black}and is calculated by rounding down the available deposited energy in the unit of $\epsilon$}. As depicted in Fig~\ref{fig:spectra}, the coherent effect significantly augments the excitation rates of the multi-phonon Migdal process both in the SM and in the presence of the neutrino magnetic moment. It should also be noted that the shape of the differential excitation spectra of multi-phonon Migdal process remains the same in the SM and in the presence of the neutrino magnetic moment, since the only energy dependent parts of Eq.~(\ref{Eq:Rate_SM}) and Eq.~(\ref{Eq:Rate_mag}) are from the last integrals of the averaged energy loss function $\mathcal{F}(\omega)$. All the other components are calculated independently and contributed as an constant.

\section{Numerical Sensitivities and Constraints}

\begin{figure*}
	\centering
	\includegraphics[width=\textwidth, angle=0]{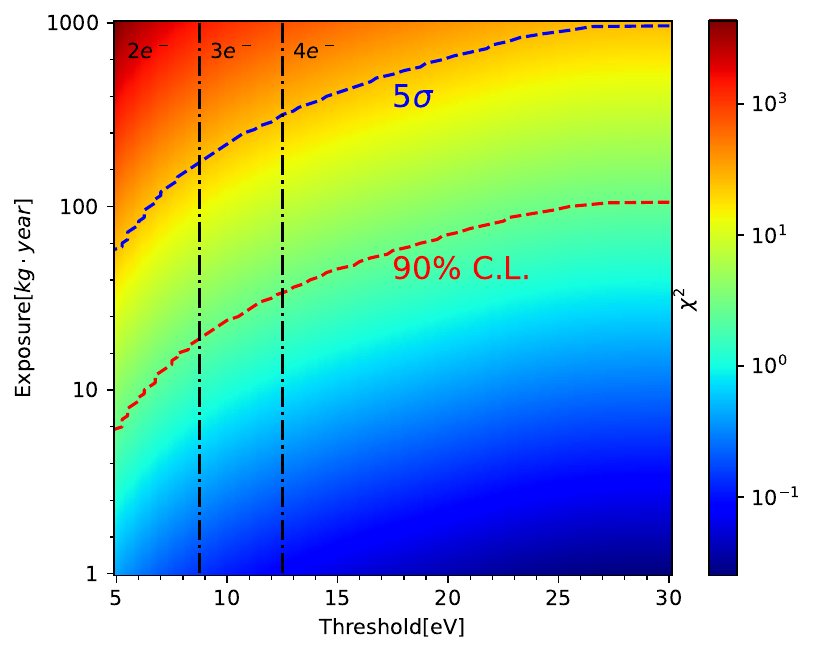}
	\caption{The sensitivities of the Migdal effect under different levels of the energy threshold and exposure. the 5$\sigma$ discovery potential and 90$\%$ C.L. limit for the Migdal effect in blue and red dash lines respectively. 
The vertical dash dotted lines are shown for the thresholds {\color{black}with different numbers of ionization electrons.}}
	\label{sen:migdal}	
\end{figure*} 
\begin{figure*}
	\centering
	\includegraphics[width=\textwidth, angle=0]{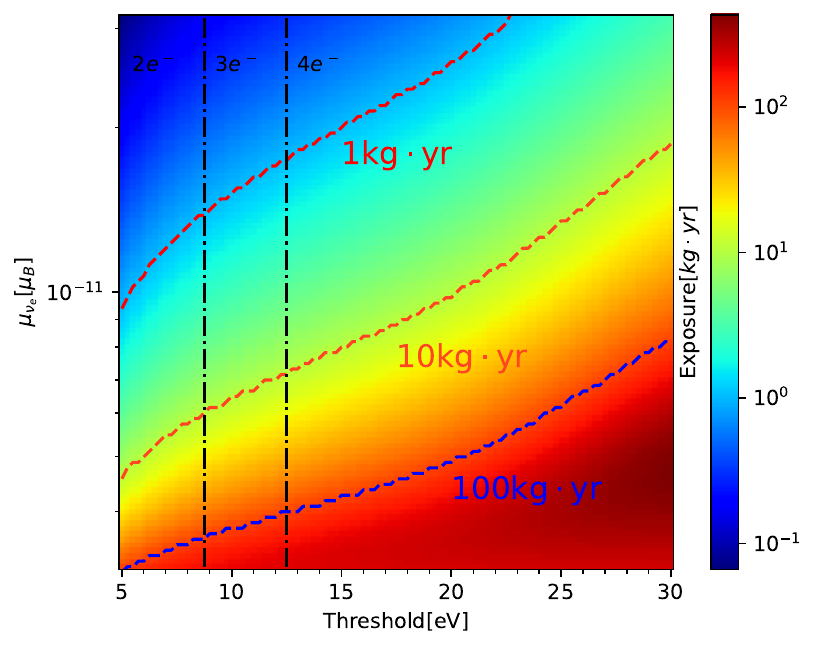}
	\caption{The necessary exposure to achieve a sensitivity level of the neutrino magnetic moment at 90$\%$ is illustrated for various energy thresholds in the presence of the Migdal effect. The contours for the exposures of 1 kg$\cdot$yr, 10 kg$\cdot$yr and 100 kg$\cdot$yr in are shown with red, orange and blue dash lines respectively. The vertical dash dotted lines are shown for the energy thresholds with {\color{black}different numbers of ionization electrons.}}
	\label{sen:numag}	
\end{figure*} 
In this section, we provide an analysis of numerical sensitivities on the Migdal effect and the corresponding constraints on the neutrino magnetic moment in the presence of the Migdal effect. We conduct our calculation based on an idealized experiment with the detector of the silicon Charge Coupled Device (CCD). 
{\color{black}It should be noted that the nuclear recoil of CE$\nu$NS without the Migdal effect does not deposit much energy in the form of excited electrons due to the quenching effect, in particular for the low energy range of interest (i.e., from a few eV to a few tens of eV). More details can be found in ref.~\cite{Knapen:2020aky,Scholz:2016qos,Essig:2018tss}. According to the ionization models in ref.~\cite{Essig:2018tss}, we find that the event rate of CE$\nu$NS without the Migdal effect is similar to that of the Migdal effect in the energy range of interest. If the CE$\nu$NS signals without the Migdal effect are taken into consideration based on the ionization models in ref.~\cite{Essig:2018tss}, they will become important backgrounds and decrease the sensitivity on the Migdal effect by a factor of $\mathcal{O}$(1-2) depending on the specific ionization model. However, the ionization models of semiconductors at such low energies are very uncertain~\cite{Essig:2018tss,Chavarria:2016xsi} and lack experimental data. Therefore, in this work we choose to neglect the signals from the CE$\nu$NS without the Migdal effect.}
{Compared with the E$\nu$ES signal, the Migdal event rate benefits from the coherent effect and the event rate is expected to be significantly enhanced, making the corresponding detection substantially easier. With these unique features, the Migdal excitation can serve as a valuable tool for the low energy phenomenology in the context of neutrino-nucleus interactions.}

In the following analysis, we restrict our attention to the events with more than two ionization signals (larger than 5 eV) and assume zero background within this energy region{, which is rather similar to the assumption and setup in ref.~\cite{Berghaus:2022pbu}}. We have excluded the events with one single ionization electron, as this energy region includes substantial amount of background signals that have not been extensively studied. 
{\color{black}For the region with more than one ionization electrons, the background electron excitations are from various sources and several comments are provided as follows:
\begin{itemize}
    \item The dark current and other electronic noises are important backgrounds on a CCD detector~\cite{Tiffenberg:2017aac} and can be dominant on some kinds of devices. However, these kinds of backgrounds depend on the designs of the devices and we will not take them into account in this work.
    \item Solar neutrinos~\cite{Davis:1968cp,Vinyoles:2018imb} and terrestrial neutrinos~\cite{Fiorentini:2003ww,Fiorentini:2005ma} are intensive natural neutrino sources at the Earth and have the similar energy range with the reactor neutrinos. However, the corresponding fluxes are much smaller than the reactor neutrinos and we have neglected them in this work.
    \item The E$\nu$ES recoils from the reactor neutrinos deposit most of their energy to excite electrons in the CCD detectors and it is hard to discriminate them from the Migdal excitation. We will consider them as the main background for the Migdal effect in this work.
    \item Atmospheric and high energy cosmic neutrinos can also induce possible backgrounds in CCD detectors despite their high energy. However, the corresponding fluxes are very low compared to reactor neutrinos and can be neglected.
    \item Since the CCD detectors are usually well shielded, the high energy signals including cosmic ray and high energy electrons and photons from the reactors are minimized. Also, the excitation traces of such signals can be easily discriminated from those of low energy neutrino induced signals~\cite{DAMIC:2015ipv}.
\end{itemize}
}

The same as in Fig~\ref{fig:spectra}, we consider one of the Taishan reactor cores with a thermal power of 4.6 $\rm GW_{\rm th}$, and a distance of 30 meters for the detector location~\cite{JUNO:2020ijm}. {To explore the projected constraints of the Migdal effect and the neutrino magnetic moment, {\color{black}we considered the Poissonian least-squares function}
{\color{black}
    \begin{equation}
	\chi^2=2\sum_{i=1}^{n}\left[(1+\epsilon_{\rm exp}) N_{i}^{\rm pred}-N_{i}^{\rm exp}+N_{i}^{\rm exp}\rm{ln}\left(\frac{N_{i}^{\rm exp}}{(1+\epsilon_{\rm exp}) N_{i}^{\rm pred}}\right)\right]+\left(\frac{\epsilon_{\rm exp}}{\sigma_{\rm exp}}\right)^{2}\,,
	\end{equation}}
{\color{black}where $N_{i}^{\rm exp}$ is the pseudo event number} of the signal of the considered experiment in the $i$th energy bin {that include the Migdal excitation and background signal}, and $N_{i}^{\rm pred}$ is the predicted event number. {In this simplified experimental setup, the reactor neutrino fluxes dominate the signals and the E$\nu$ES excitation is the main background source for the Migdal excitation. We tentatively neglect other important radioactive and cosmogenic backgrounds for further detailed studies.} $\epsilon_{\rm exp}$ is the simplified nuisance parameter which quantifies the total detection uncertainty with $\sigma_{\rm exp}=5\%$ being the  expected standard deviation.
The current semiconductor detectors are extremely sensitive to electron ionization signals and it is a convenient approximation to employ an ideal efficiency of 100\% for illustration.}

In Fig.~\ref{sen:migdal}, we illustrate the sensitivities of the Migdal effect under different levels of the energy threshold and exposure. We present the 5$\sigma$ discovery potential and 90$\%$ confidence level (C.L.) limit for the Migdal effect in blue and red dash lines respectively. 
The vertical dash dotted lines are shown for the thresholds to observe {\color{black}different numbers of ionization electrons.}
One can observe that with a energy threshold of 8.8 eV, which enables us to detect all events with more than two ionization signals, we need an exposure of few {\color{black}tens of} kg$\cdot$yr to reach the limit of 90$\%$ C.L., while an exposure of a few {\color{black}hundreds} of kg$\cdot$yr is needed to confirm the discovery at 5$\sigma$.
Thus for the present detector technology with less than one to a few kg$\cdot$yr of exposure~\cite{SENSEI:2020dpa,SuperCDMS:2018mne} {\color{black}there will not be enough statistics to confirm the existence of the Migdal effect.} In the future, the silicon based CCD detectors could reach the exposure level of several tens of kg$\cdot$yr or even one hundred {\color{black}while still keep the capability to detect single electron excitations in pixels}, which will give more opportunity to probe the current prediction of the Migdal effect. 
A better sensitivity can also be achieved by lowering the threshold of ionization signals, but in this case the requirement on the CCD readout and background filtering would be more challenging~\cite{EDELWEISS:2017uga,EDELWEISS:2020fxc,EDELWEISS:2022ktt}.

In Fig.~\ref{sen:numag} the necessary exposure to achieve a sensitivity level of the neutrino magnetic moment at 90$\%$ is illustrated for various energy thresholds in the presence of the Migdal effect. 
We illustrate the contours for the exposures of 1 kg$\cdot$yr, 10 kg$\cdot$yr and 100 kg$\cdot$yr in red, orange and blue dash lines respectively.
The vertical dash dotted lines are shown for the thresholds to observe {\color{black}different numbers of ionization electrons}.
Notably, with the contribution of the Migdal effect, a silicon-based detector with an exposure of 1 kg$\cdot$yr and {\color{black}capability to read out all two electron} ionization signals (above 5 eV), can readily reach the constraint level that is comparable to those of LUZ-ZEPLIN~\cite{AtzoriCorona:2022jeb,Giunti:2023yha} or XENON1T~\cite{XENON:2020rca}.
Moreover, we also observe from Fig.~\ref{sen:numag} that reducing the threshold form 20 eV to 5 eV is equivalent to increasing the exposure by approximately ten fold, which implies the necessity of detecting the low energy events in order to fully utilize the potential of detectors.


\section{Conclusion} 

In this work we have presented for the first time a framework of the phonon-mediated Migdal effect of the neutrino nucleus scattering in the semiconductors.
We have calculated the neutrino-induced Migdal excitation rates in semiconductors with the multi-phonon model in the SM and in the presence of the neutrino magnetic moment. We have shown that the contribution from the Migdal effect is dramatically enhanced by the coherent effect of the neutrino nucleus scattering.
Meanwhile, the Migdal excitation can be distinguished from the CE$\nu$ES process without the Migdal effect by observing the presence of additional electron excitations. Consequently, these excellent features make the Migdal excitation an effective tool to investigate the new physics phenomena at low energy scales.

Utilizing the new theoretical calculation, we have performed a numerical simulation on the sensitivities of the Migdal effect and neutrino magnetic moment with reactor neutrinos and presented the detection potential under different detection energy thresholds and exposures, which demonstrate the excellent prospect for testing the Migdal effect in reactor neutrino experiments and further provide a unique frontier to probe the new physics beyond the SM.


\acknowledgments
The authors are grateful to Dr.~Zhengliang Liang for helpful discussions. This work was supported in part by National Natural Science Foundation of China under Grant Nos.~12075255, 12075254 and 11835013.

\bibliographystyle{JHEP}
\bibliography{main}

\providecommand{\href}[2]{#2}\begingroup\raggedright\begin{thebibliography}{10}

\bibitem{SuperCDMS:2018mne}
{\scshape SuperCDMS} collaboration, \emph{{First Dark Matter Constraints from a
  SuperCDMS Single-Charge Sensitive Detector}},
  \href{https://doi.org/10.1103/PhysRevLett.121.051301}{\emph{Phys. Rev. Lett.}
  {\bfseries 121} (2018) 051301}
  [\href{https://arxiv.org/abs/1804.10697}{{\ttfamily 1804.10697}}].

\bibitem{EDELWEISS:2020fxc}
{\scshape EDELWEISS} collaboration, \emph{{First germanium-based constraints on
  sub-MeV Dark Matter with the EDELWEISS experiment}},
  \href{https://doi.org/10.1103/PhysRevLett.125.141301}{\emph{Phys. Rev. Lett.}
  {\bfseries 125} (2020) 141301}
  [\href{https://arxiv.org/abs/2003.01046}{{\ttfamily 2003.01046}}].

\bibitem{SENSEI:2020dpa}
{\scshape SENSEI} collaboration, \emph{{SENSEI: Direct-Detection Results on
  sub-GeV Dark Matter from a New Skipper-CCD}},
  \href{https://doi.org/10.1103/PhysRevLett.125.171802}{\emph{Phys. Rev. Lett.}
  {\bfseries 125} (2020) 171802}
  [\href{https://arxiv.org/abs/2004.11378}{{\ttfamily 2004.11378}}].

\bibitem{LZ:2022ufs}
{\scshape LZ} collaboration, \emph{{First Dark Matter Search Results from the
  LUX-ZEPLIN (LZ) Experiment}},
  \href{https://arxiv.org/abs/2207.03764}{{\ttfamily 2207.03764}}.

\bibitem{XENON:2018voc}
{\scshape XENON} collaboration, \emph{{Dark Matter Search Results from a One
  Ton-Year Exposure of XENON1T}},
  \href{https://doi.org/10.1103/PhysRevLett.121.111302}{\emph{Phys. Rev. Lett.}
  {\bfseries 121} (2018) 111302}
  [\href{https://arxiv.org/abs/1805.12562}{{\ttfamily 1805.12562}}].

\bibitem{XENON:2020fgj}
{\scshape XENON} collaboration, \emph{{Search for inelastic scattering of WIMP
  dark matter in XENON1T}},
  \href{https://doi.org/10.1103/PhysRevD.103.063028}{\emph{Phys. Rev. D}
  {\bfseries 103} (2021) 063028}
  [\href{https://arxiv.org/abs/2011.10431}{{\ttfamily 2011.10431}}].

\bibitem{Liang:2021zkg}
Z.-L.~Liang, C.~Mo and P.~Zhang, \emph{{In-medium screening effects for the
  Galactic halo and solar-reflected dark matter detection in semiconductor
  targets}}, \href{https://doi.org/10.1103/PhysRevD.104.096001}{\emph{Phys.
  Rev. D} {\bfseries 104} (2021) 096001}
  [\href{https://arxiv.org/abs/2107.01209}{{\ttfamily 2107.01209}}].

\bibitem{Liang:2018bdb}
Z.-L.~Liang, L.~Zhang, P.~Zhang and F.~Zheng, \emph{{The wavefunction
  reconstruction effects in calculation of DM-induced electronic transition in
  semiconductor targets}},
  \href{https://doi.org/10.1007/JHEP01(2019)149}{\emph{JHEP} {\bfseries 01}
  (2019) 149} [\href{https://arxiv.org/abs/1810.13394}{{\ttfamily
  1810.13394}}].

\bibitem{Graham:2012su}
P.W.~Graham, D.E.~Kaplan, S.~Rajendran and M.T.~Walters, \emph{{Semiconductor
  Probes of Light Dark Matter}},
  \href{https://doi.org/10.1016/j.dark.2012.09.001}{\emph{Phys. Dark Univ.}
  {\bfseries 1} (2012) 32} [\href{https://arxiv.org/abs/1203.2531}{{\ttfamily
  1203.2531}}].

\bibitem{Essig:2015cda}
R.~Essig, M.~Fernandez-Serra, J.~Mardon, A.~Soto, T.~Volansky and T.-T.~Yu,
  \emph{{Direct Detection of sub-GeV Dark Matter with Semiconductor Targets}},
  \href{https://doi.org/10.1007/JHEP05(2016)046}{\emph{JHEP} {\bfseries 05}
  (2016) 046} [\href{https://arxiv.org/abs/1509.01598}{{\ttfamily
  1509.01598}}].

\bibitem{Hochberg:2016sqx}
Y.~Hochberg, T.~Lin and K.M.~Zurek, \emph{{Absorption of light dark matter in
  semiconductors}},
  \href{https://doi.org/10.1103/PhysRevD.95.023013}{\emph{Phys. Rev. D}
  {\bfseries 95} (2017) 023013}
  [\href{https://arxiv.org/abs/1608.01994}{{\ttfamily 1608.01994}}].

\bibitem{Caputo:2019xum}
A.~Caputo, A.~Esposito, E.~Geoffray, A.D.~Polosa and S.~Sun, \emph{{Dark
  Matter, Dark Photon and Superfluid He-4 from Effective Field Theory}},
  \href{https://doi.org/10.1016/j.physletb.2020.135258}{\emph{Phys. Lett. B}
  {\bfseries 802} (2020) 135258}
  [\href{https://arxiv.org/abs/1911.04511}{{\ttfamily 1911.04511}}].

\bibitem{Caputo:2019cyg}
A.~Caputo, A.~Esposito and A.D.~Polosa, \emph{{Sub-MeV Dark Matter and the
  Goldstone Modes of Superfluid Helium}},
  \href{https://doi.org/10.1103/PhysRevD.100.116007}{\emph{Phys. Rev. D}
  {\bfseries 100} (2019) 116007}
  [\href{https://arxiv.org/abs/1907.10635}{{\ttfamily 1907.10635}}].

\bibitem{Knapen:2016cue}
S.~Knapen, T.~Lin and K.M.~Zurek, \emph{{Light Dark Matter in Superfluid
  Helium: Detection with Multi-excitation Production}},
  \href{https://doi.org/10.1103/PhysRevD.95.056019}{\emph{Phys. Rev. D}
  {\bfseries 95} (2017) 056019}
  [\href{https://arxiv.org/abs/1611.06228}{{\ttfamily 1611.06228}}].

\bibitem{Hochberg:2015pha}
Y.~Hochberg, Y.~Zhao and K.M.~Zurek, \emph{{Superconducting Detectors for
  Superlight Dark Matter}},
  \href{https://doi.org/10.1103/PhysRevLett.116.011301}{\emph{Phys. Rev. Lett.}
  {\bfseries 116} (2016) 011301}
  [\href{https://arxiv.org/abs/1504.07237}{{\ttfamily 1504.07237}}].

\bibitem{Hochberg:2016ajh}
Y.~Hochberg, T.~Lin and K.M.~Zurek, \emph{{Detecting Ultralight Bosonic Dark
  Matter via Absorption in Superconductors}},
  \href{https://doi.org/10.1103/PhysRevD.94.015019}{\emph{Phys. Rev. D}
  {\bfseries 94} (2016) 015019}
  [\href{https://arxiv.org/abs/1604.06800}{{\ttfamily 1604.06800}}].

\bibitem{Papoulias:2017qdn}
D.K.~Papoulias and T.S.~Kosmas, \emph{{COHERENT constraints to conventional and
  exotic neutrino physics}},
  \href{https://doi.org/10.1103/PhysRevD.97.033003}{\emph{Phys. Rev. D}
  {\bfseries 97} (2018) 033003}
  [\href{https://arxiv.org/abs/1711.09773}{{\ttfamily 1711.09773}}].

\bibitem{Cadeddu:2018izq}
M.~Cadeddu and F.~Dordei, \emph{{Reinterpreting the weak mixing angle from
  atomic parity violation in view of the Cs neutron rms radius measurement from
  COHERENT}}, \href{https://doi.org/10.1103/PhysRevD.99.033010}{\emph{Phys.
  Rev. D} {\bfseries 99} (2019) 033010}
  [\href{https://arxiv.org/abs/1808.10202}{{\ttfamily 1808.10202}}].

\bibitem{Cadeddu:2018dux}
M.~Cadeddu, C.~Giunti, K.A.~Kouzakov, Y.F.~Li, A.I.~Studenikin and Y.Y.~Zhang,
  \emph{{Neutrino Charge Radii from COHERENT Elastic Neutrino-Nucleus
  Scattering}}, \href{https://doi.org/10.1103/PhysRevD.98.113010}{\emph{Phys.
  Rev. D} {\bfseries 98} (2018) 113010}
  [\href{https://arxiv.org/abs/1810.05606}{{\ttfamily 1810.05606}}].

\bibitem{Cadeddu:2019eta}
M.~Cadeddu, F.~Dordei, C.~Giunti, Y.F.~Li and Y.Y.~Zhang, \emph{{Neutrino,
  electroweak, and nuclear physics from COHERENT elastic neutrino-nucleus
  scattering with refined quenching factor}},
  \href{https://doi.org/10.1103/PhysRevD.101.033004}{\emph{Phys. Rev. D}
  {\bfseries 101} (2020) 033004}
  [\href{https://arxiv.org/abs/1908.06045}{{\ttfamily 1908.06045}}].

\bibitem{Kim:2021lun}
J.E.~Kim, A.~Dasgupta and S.K.~Kang, \emph{{Probing Neutrino Dipole Portal at
  COHERENT Experiment}},  \href{https://arxiv.org/abs/2108.12998}{{\ttfamily
  2108.12998}}.

\bibitem{Coloma:2017ncl}
P.~Coloma, M.C.~Gonzalez-Garcia, M.~Maltoni and T.~Schwetz, \emph{{COHERENT
  Enlightenment of the Neutrino Dark Side}},
  \href{https://doi.org/10.1103/PhysRevD.96.115007}{\emph{Phys. Rev. D}
  {\bfseries 96} (2017) 115007}
  [\href{https://arxiv.org/abs/1708.02899}{{\ttfamily 1708.02899}}].

\bibitem{Liao:2017uzy}
J.~Liao and D.~Marfatia, \emph{{COHERENT constraints on nonstandard neutrino
  interactions}},
  \href{https://doi.org/10.1016/j.physletb.2017.10.046}{\emph{Phys. Lett. B}
  {\bfseries 775} (2017) 54}
  [\href{https://arxiv.org/abs/1708.04255}{{\ttfamily 1708.04255}}].

\bibitem{Denton:2018xmq}
P.B.~Denton, Y.~Farzan and I.M.~Shoemaker, \emph{{Testing large non-standard
  neutrino interactions with arbitrary mediator mass after COHERENT data}},
  \href{https://doi.org/10.1007/JHEP07(2018)037}{\emph{JHEP} {\bfseries 07}
  (2018) 037} [\href{https://arxiv.org/abs/1804.03660}{{\ttfamily
  1804.03660}}].

\bibitem{AristizabalSierra:2018eqm}
D.~Aristizabal~Sierra, V.~De~Romeri and N.~Rojas, \emph{{COHERENT analysis of
  neutrino generalized interactions}},
  \href{https://doi.org/10.1103/PhysRevD.98.075018}{\emph{Phys. Rev. D}
  {\bfseries 98} (2018) 075018}
  [\href{https://arxiv.org/abs/1806.07424}{{\ttfamily 1806.07424}}].

\bibitem{Giunti:2019xpr}
C.~Giunti, \emph{{General COHERENT constraints on neutrino nonstandard
  interactions}},
  \href{https://doi.org/10.1103/PhysRevD.101.035039}{\emph{Phys. Rev. D}
  {\bfseries 101} (2020) 035039}
  [\href{https://arxiv.org/abs/1909.00466}{{\ttfamily 1909.00466}}].

\bibitem{EDELWEISS:2017uga}
{\scshape EDELWEISS} collaboration, \emph{{Optimizing EDELWEISS detectors for
  low-mass WIMP searches}},
  \href{https://doi.org/10.1103/PhysRevD.97.022003}{\emph{Phys. Rev. D}
  {\bfseries 97} (2018) 022003}
  [\href{https://arxiv.org/abs/1707.04308}{{\ttfamily 1707.04308}}].

\bibitem{migdal:1939svj}
A.~Migdal, \emph{{Ionizatsiya atomov pri yadernykh reaktsiyakh}}, {\emph{Sov.
  Phys. JETP} {\bfseries 9} (1939) 1163}.

\bibitem{Ibe:2017yqa}
M.~Ibe, W.~Nakano, Y.~Shoji and K.~Suzuki, \emph{{Migdal Effect in Dark Matter
  Direct Detection Experiments}},
  \href{https://doi.org/10.1007/JHEP03(2018)194}{\emph{JHEP} {\bfseries 03}
  (2018) 194} [\href{https://arxiv.org/abs/1707.07258}{{\ttfamily
  1707.07258}}].

\bibitem{Essig:2019xkx}
R.~Essig, J.~Pradler, M.~Sholapurkar and T.-T.~Yu, \emph{{Relation between the
  Migdal Effect and Dark Matter-Electron Scattering in Isolated Atoms and
  Semiconductors}},
  \href{https://doi.org/10.1103/PhysRevLett.124.021801}{\emph{Phys. Rev. Lett.}
  {\bfseries 124} (2020) 021801}
  [\href{https://arxiv.org/abs/1908.10881}{{\ttfamily 1908.10881}}].

\bibitem{Bell:2019egg}
N.F.~Bell, J.B.~Dent, J.L.~Newstead, S.~Sabharwal and T.J.~Weiler,
  \emph{{Migdal effect and photon bremsstrahlung in effective field theories of
  dark matter direct detection and coherent elastic neutrino-nucleus
  scattering}}, \href{https://doi.org/10.1103/PhysRevD.101.015012}{\emph{Phys.
  Rev. D} {\bfseries 101} (2020) 015012}
  [\href{https://arxiv.org/abs/1905.00046}{{\ttfamily 1905.00046}}].

\bibitem{Dolan:2017xbu}
M.J.~Dolan, F.~Kahlhoefer and C.~McCabe, \emph{{Directly detecting sub-GeV dark
  matter with electrons from nuclear scattering}},
  \href{https://doi.org/10.1103/PhysRevLett.121.101801}{\emph{Phys. Rev. Lett.}
  {\bfseries 121} (2018) 101801}
  [\href{https://arxiv.org/abs/1711.09906}{{\ttfamily 1711.09906}}].

\bibitem{Berghaus:2022pbu}
K.V.~Berghaus, A.~Esposito, R.~Essig and M.~Sholapurkar, \emph{{The Migdal
  effect in semiconductors for dark matter with masses below \ensuremath{\sim}
  100 MeV}}, \href{https://doi.org/10.1007/JHEP01(2023)023}{\emph{JHEP}
  {\bfseries 01} (2023) 023}
  [\href{https://arxiv.org/abs/2210.06490}{{\ttfamily 2210.06490}}].

\bibitem{Gelmini:2020xir}
G.B.~Gelmini, V.~Takhistov and E.~Vitagliano, \emph{{Scalar direct detection:
  In-medium effects}},
  \href{https://doi.org/10.1016/j.physletb.2020.135779}{\emph{Phys. Lett. B}
  {\bfseries 809} (2020) 135779}
  [\href{https://arxiv.org/abs/2006.13909}{{\ttfamily 2006.13909}}].

\bibitem{Knapen:2020aky}
S.~Knapen, J.~Kozaczuk and T.~Lin, \emph{{Migdal Effect in Semiconductors}},
  \href{https://doi.org/10.1103/PhysRevLett.127.081805}{\emph{Phys. Rev. Lett.}
  {\bfseries 127} (2021) 081805}
  [\href{https://arxiv.org/abs/2011.09496}{{\ttfamily 2011.09496}}].

\bibitem{Liang:2020ryg}
Z.-L.~Liang, C.~Mo, F.~Zheng and P.~Zhang, \emph{{Describing the Migdal effect
  with a bremsstrahlung-like process and many-body effects}},
  \href{https://doi.org/10.1103/PhysRevD.104.056009}{\emph{Phys. Rev. D}
  {\bfseries 104} (2021) 056009}
  [\href{https://arxiv.org/abs/2011.13352}{{\ttfamily 2011.13352}}].

\bibitem{Liang:2019nnx}
Z.-L.~Liang, L.~Zhang, F.~Zheng and P.~Zhang, \emph{{Describing Migdal effects
  in diamond crystal with atom-centered localized Wannier functions}},
  \href{https://doi.org/10.1103/PhysRevD.102.043007}{\emph{Phys. Rev. D}
  {\bfseries 102} (2020) 043007}
  [\href{https://arxiv.org/abs/1912.13484}{{\ttfamily 1912.13484}}].

\bibitem{Liang:2022xbu}
Z.-L.~Liang, C.~Mo, F.~Zheng and P.~Zhang, \emph{{Phonon-mediated Migdal effect
  in semiconductor detectors}},
  \href{https://doi.org/10.1103/PhysRevD.106.043004}{\emph{Phys. Rev. D}
  {\bfseries 106} (2022) 043004}
  [\href{https://arxiv.org/abs/2205.03395}{{\ttfamily 2205.03395}}].

\bibitem{Bell:2021ihi}
N.F.~Bell, J.B.~Dent, R.F.~Lang, J.L.~Newstead and A.C.~Ritter,
  \emph{{Observing the Migdal effect from nuclear recoils of neutral particles
  with liquid xenon and argon detectors}},
  \href{https://doi.org/10.1103/PhysRevD.105.096015}{\emph{Phys. Rev. D}
  {\bfseries 105} (2022) 096015}
  [\href{https://arxiv.org/abs/2112.08514}{{\ttfamily 2112.08514}}].

\bibitem{Araujo:2022wjh}
H.M.~Ara\'ujo et~al., \emph{{The MIGDAL experiment: Measuring a rare atomic
  process to aid the search for dark matter}},
  \href{https://arxiv.org/abs/2207.08284}{{\ttfamily 2207.08284}}.

\bibitem{Cox:2022ekg}
P.~Cox, M.J.~Dolan, C.~McCabe and H.M.~Quiney, \emph{{Precise predictions and
  new insights for atomic ionisation from the Migdal effect}},
  \href{https://arxiv.org/abs/2208.12222}{{\ttfamily 2208.12222}}.

\bibitem{Adams:2022zvg}
D.~Adams, D.~Baxter, H.~Day, R.~Essig and Y.~Kahn, \emph{{Measuring the Migdal
  effect in semiconductors for dark matter detection}},
  \href{https://doi.org/10.1103/PhysRevD.107.L041303}{\emph{Phys. Rev. D}
  {\bfseries 107} (2023) L041303}
  [\href{https://arxiv.org/abs/2210.04917}{{\ttfamily 2210.04917}}].

\bibitem{Xu:2023wev}
J.~Xu et~al., \emph{{Search for the Migdal effect in liquid xenon with
  keV-level nuclear recoils}},
  \href{https://arxiv.org/abs/2307.12952}{{\ttfamily 2307.12952}}.

\bibitem{EDELWEISS:2022ktt}
{\scshape EDELWEISS} collaboration, \emph{{Search for sub-GeV dark matter via
  the Migdal effect with an EDELWEISS germanium detector with NbSi
  transition-edge sensors}},
  \href{https://doi.org/10.1103/PhysRevD.106.062004}{\emph{Phys. Rev. D}
  {\bfseries 106} (2022) 062004}
  [\href{https://arxiv.org/abs/2203.03993}{{\ttfamily 2203.03993}}].

\bibitem{LUX:2018akb}
{\scshape LUX} collaboration, \emph{{Results of a Search for Sub-GeV Dark
  Matter Using 2013 LUX Data}},
  \href{https://doi.org/10.1103/PhysRevLett.122.131301}{\emph{Phys. Rev. Lett.}
  {\bfseries 122} (2019) 131301}
  [\href{https://arxiv.org/abs/1811.11241}{{\ttfamily 1811.11241}}].

\bibitem{CDEX:2019hzn}
{\scshape CDEX} collaboration, \emph{{Constraints on Spin-Independent Nucleus
  Scattering with sub-GeV Weakly Interacting Massive Particle Dark Matter from
  the CDEX-1B Experiment at the China Jinping Underground Laboratory}},
  \href{https://doi.org/10.1103/PhysRevLett.123.161301}{\emph{Phys. Rev. Lett.}
  {\bfseries 123} (2019) 161301}
  [\href{https://arxiv.org/abs/1905.00354}{{\ttfamily 1905.00354}}].

\bibitem{EDELWEISS:2019vjv}
{\scshape EDELWEISS} collaboration, \emph{{Searching for low-mass dark matter
  particles with a massive Ge bolometer operated above-ground}},
  \href{https://doi.org/10.1103/PhysRevD.99.082003}{\emph{Phys. Rev. D}
  {\bfseries 99} (2019) 082003}
  [\href{https://arxiv.org/abs/1901.03588}{{\ttfamily 1901.03588}}].

\bibitem{XENON:2019zpr}
{\scshape XENON} collaboration, \emph{{Search for Light Dark Matter
  Interactions Enhanced by the Migdal Effect or Bremsstrahlung in XENON1T}},
  \href{https://doi.org/10.1103/PhysRevLett.123.241803}{\emph{Phys. Rev. Lett.}
  {\bfseries 123} (2019) 241803}
  [\href{https://arxiv.org/abs/1907.12771}{{\ttfamily 1907.12771}}].

\bibitem{Essig:2022dfa}
R.~Essig, G.K.~Giovanetti, N.~Kurinsky, D.~McKinsey, K.~Ramanathan, K.~Stifter
  et~al., \emph{{Snowmass2021 Cosmic Frontier: The landscape of low-threshold
  dark matter direct detection in the next decade}},  in \emph{{2022 Snowmass
  Summer Study}}, 3, 2022 [\href{https://arxiv.org/abs/2203.08297}{{\ttfamily
  2203.08297}}].

\bibitem{Drukier:1984vhf}
A.~Drukier and L.~Stodolsky, \emph{{Principles and Applications of a Neutral
  Current Detector for Neutrino Physics and Astronomy}},
  \href{https://doi.org/10.1103/PhysRevD.30.2295}{\emph{Phys. Rev. D}
  {\bfseries 30} (1984) 2295}.

\bibitem{Patton:2012jr}
K.~Patton, J.~Engel, G.C.~McLaughlin and N.~Schunck, \emph{{Neutrino-nucleus
  coherent scattering as a probe of neutron density distributions}},
  \href{https://doi.org/10.1103/PhysRevC.86.024612}{\emph{Phys. Rev. C}
  {\bfseries 86} (2012) 024612}
  [\href{https://arxiv.org/abs/1207.0693}{{\ttfamily 1207.0693}}].

\bibitem{Tiffenberg:2017aac}
{\scshape SENSEI} collaboration, \emph{{Single-electron and single-photon
  sensitivity with a silicon Skipper CCD}},
  \href{https://doi.org/10.1103/PhysRevLett.119.131802}{\emph{Phys. Rev. Lett.}
  {\bfseries 119} (2017) 131802}
  [\href{https://arxiv.org/abs/1706.00028}{{\ttfamily 1706.00028}}].

\bibitem{PhysRevLett.45.963}
K.~Fujikawa and R.E.~Shrock, \emph{Magnetic moment of a massive neutrino and
  neutrino-spin rotation},
  \href{https://doi.org/10.1103/PhysRevLett.45.963}{\emph{Phys. Rev. Lett.}
  {\bfseries 45} (1980) 963}.

\bibitem{Giunti:2014ixa}
C.~Giunti and A.~Studenikin, \emph{{Neutrino electromagnetic interactions: a
  window to new physics}},
  \href{https://doi.org/10.1103/RevModPhys.87.531}{\emph{Rev. Mod. Phys.}
  {\bfseries 87} (2015) 531} [\href{https://arxiv.org/abs/1403.6344}{{\ttfamily
  1403.6344}}].

\bibitem{Bell:2006wi}
N.F.~Bell, M.~Gorchtein, M.J.~Ramsey-Musolf, P.~Vogel and P.~Wang, \emph{{Model
  independent bounds on magnetic moments of Majorana neutrinos}},
  \href{https://doi.org/10.1016/j.physletb.2006.09.055}{\emph{Phys. Lett. B}
  {\bfseries 642} (2006) 377}
  [\href{https://arxiv.org/abs/hep-ph/0606248}{{\ttfamily hep-ph/0606248}}].

\bibitem{Bell:2005kz}
N.F.~Bell, V.~Cirigliano, M.J.~Ramsey-Musolf, P.~Vogel and M.B.~Wise,
  \emph{{How magnetic is the Dirac neutrino?}},
  \href{https://doi.org/10.1103/PhysRevLett.95.151802}{\emph{Phys. Rev. Lett.}
  {\bfseries 95} (2005) 151802}
  [\href{https://arxiv.org/abs/hep-ph/0504134}{{\ttfamily hep-ph/0504134}}].

\bibitem{Amsler:2002tu}
{\scshape MUNU} collaboration, \emph{A new measurement of the anti-nu/e e-
  elastic cross section at very low energy}, {\emph{Phys. Lett.} {\bfseries
  B545} (2002) 57}.

\bibitem{TEXONO:2002pra}
{\scshape TEXONO} collaboration, \emph{{Limit on the electron neutrino magnetic
  moment from the Kuo-Sheng reactor neutrino experiment}},
  \href{https://doi.org/10.1103/PhysRevLett.90.131802}{\emph{Phys. Rev. Lett.}
  {\bfseries 90} (2003) 131802}
  [\href{https://arxiv.org/abs/hep-ex/0212003}{{\ttfamily hep-ex/0212003}}].

\bibitem{Beda:2007hf}
{\scshape GEMMA} collaboration, \emph{The first result of the neutrino magnetic
  moment measurement in the gemma experiment}, {\emph{Phys. Atom. Nucl.}
  {\bfseries 70} (2007) 1873}
  [\href{https://arxiv.org/abs/0705.4576}{{\ttfamily 0705.4576}}].

\bibitem{PhysRevLett.111.231301}
N.~Viaux, M.~Catelan, P.B.~Stetson, G.G.~Raffelt, J.~Redondo, A.A.R.~Valcarce
  et~al., \emph{Neutrino and axion bounds from the globular cluster m5 (ngc
  5904)}, \href{https://doi.org/10.1103/PhysRevLett.111.231301}{\emph{Phys.
  Rev. Lett.} {\bfseries 111} (2013) 231301}.

\bibitem{C_rsico_2014}
A.~C{\'{o}}rsico, L.~Althaus, M.M.~Bertolami, S.~Kepler and
  E.~Garc{\'{\i}}a-Berro, \emph{Constraining the neutrino magnetic dipole
  moment from white dwarf pulsations},
  \href{https://doi.org/10.1088/1475-7516/2014/08/054}{\emph{Journal of
  Cosmology and Astroparticle Physics} {\bfseries 2014} (2014) 054}.

\bibitem{Carenza:2022ngg}
P.~Carenza, G.~Lucente, M.~Gerbino, M.~Giannotti and M.~Lattanzi, \emph{{Strong
  cosmological constraints on the neutrino magnetic moment}},
  \href{https://arxiv.org/abs/2211.10432}{{\ttfamily 2211.10432}}.

\bibitem{Li:2022dkc}
S.-P.~Li and X.-J.~Xu, \emph{{Neutrino magnetic moments meet precision
  N$_{eff}$ measurements}},
  \href{https://doi.org/10.1007/JHEP02(2023)085}{\emph{JHEP} {\bfseries 02}
  (2023) 085} [\href{https://arxiv.org/abs/2211.04669}{{\ttfamily
  2211.04669}}].

\bibitem{Li:2023jhep}
Y.-F.~Li and S.-y.~Xia, \emph{{Enhancement of the screening effect in
  semiconductor detectors in the presence of the neutrino magnetic moment}},
  \href{https://doi.org/10.1007/JHEP10(2023)021}{\emph{Journal of High Energy
  Physics} {\bfseries 10} (2023) 21}.

\bibitem{Debye:1913adp}
P.~Debye, \emph{Interferenz von röntgenstrahlen und wärmebewegung},
  \href{https://doi.org/https://doi.org/10.1002/andp.19133480105}{\emph{Annalen
  der Physik} {\bfseries 348} (1913) 49}
  [\href{https://arxiv.org/abs/https://onlinelibrary.wiley.com/doi/pdf/10.1002/andp.19133480105}{{\ttfamily
  https://onlinelibrary.wiley.com/doi/pdf/10.1002/andp.19133480105}}].

\bibitem{Waller:1923zfp}
I.~Waller, \emph{Zur frage der einwirkung der wärmebewegung auf die
  interferenz von röntgenstrahlen},
  \href{https://doi.org/https://doi.org/10.1007/BF01328696}{\emph{Zeitschrift
  für Physik} {\bfseries 17} (1923) 398}.

\bibitem{JUNO:2020ijm}
{\scshape JUNO} collaboration, \emph{{TAO Conceptual Design Report: A Precision
  Measurement of the Reactor Antineutrino Spectrum with Sub-percent Energy
  Resolution}},  \href{https://arxiv.org/abs/2005.08745}{{\ttfamily
  2005.08745}}.

\bibitem{Scholz:2016qos}
B.J.~Scholz, A.E.~Chavarria, J.I.~Collar, P.~Privitera and A.E.~Robinson,
  \emph{{Measurement of the low-energy quenching factor in germanium using an
  $^{88}$Y/Be photoneutron source}},
  \href{https://doi.org/10.1103/PhysRevD.94.122003}{\emph{Phys. Rev. D}
  {\bfseries 94} (2016) 122003}
  [\href{https://arxiv.org/abs/1608.03588}{{\ttfamily 1608.03588}}].

\bibitem{Essig:2018tss}
R.~Essig, M.~Sholapurkar and T.-T.~Yu, \emph{{Solar Neutrinos as a Signal and
  Background in Direct-Detection Experiments Searching for Sub-GeV Dark Matter
  With Electron Recoils}},
  \href{https://doi.org/10.1103/PhysRevD.97.095029}{\emph{Phys. Rev. D}
  {\bfseries 97} (2018) 095029}
  [\href{https://arxiv.org/abs/1801.10159}{{\ttfamily 1801.10159}}].

\bibitem{Chavarria:2016xsi}
A.E.~Chavarria et~al., \emph{{Measurement of the ionization produced by sub-keV
  silicon nuclear recoils in a CCD dark matter detector}},
  \href{https://doi.org/10.1103/PhysRevD.94.082007}{\emph{Phys. Rev. D}
  {\bfseries 94} (2016) 082007}
  [\href{https://arxiv.org/abs/1608.00957}{{\ttfamily 1608.00957}}].

\bibitem{Davis:1968cp}
R.~Davis, Jr., D.S.~Harmer and K.C.~Hoffman, \emph{{Search for neutrinos from
  the sun}}, \href{https://doi.org/10.1103/PhysRevLett.20.1205}{\emph{Phys.
  Rev. Lett.} {\bfseries 20} (1968) 1205}.

\bibitem{Vinyoles:2018imb}
N.~Vinyoles, A.~Serenelli and F.L.~Villante, \emph{{The B16 Standard Solar
  Models}}, \href{https://doi.org/10.1088/1742-6596/1056/1/012058}{\emph{J.
  Phys. Conf. Ser.} {\bfseries 1056} (2018) 012058}.

\bibitem{Fiorentini:2003ww}
G.~Fiorentini, T.~Lasserre, M.~Lissia, B.~Ricci and S.~Schonert,
  \emph{{KamLAND, terrestrial heat sources and neutrino oscillations}},
  \href{https://doi.org/10.1016/S0370-2693(03)00240-5}{\emph{Phys. Lett. B}
  {\bfseries 558} (2003) 15}
  [\href{https://arxiv.org/abs/hep-ph/0301042}{{\ttfamily hep-ph/0301042}}].

\bibitem{Fiorentini:2005ma}
G.~Fiorentini, M.~Lissia, F.~Mantovani and B.~Ricci, \emph{{KamLAND results and
  the radiogenic terrestrial heat}},
  \href{https://doi.org/10.1016/j.physletb.2005.09.067}{\emph{Phys. Lett. B}
  {\bfseries 629} (2005) 77}
  [\href{https://arxiv.org/abs/hep-ph/0508048}{{\ttfamily hep-ph/0508048}}].

\bibitem{DAMIC:2015ipv}
{\scshape DAMIC} collaboration, \emph{{Measurement of radioactive contamination
  in the high-resistivity silicon CCDs of the DAMIC experiment}},
  \href{https://doi.org/10.1088/1748-0221/10/08/P08014}{\emph{JINST} {\bfseries
  10} (2015) P08014} [\href{https://arxiv.org/abs/1506.02562}{{\ttfamily
  1506.02562}}].

\bibitem{AtzoriCorona:2022jeb}
M.~Atzori~Corona, W.M.~Bonivento, M.~Cadeddu, N.~Cargioli and F.~Dordei,
  \emph{{New constraint on neutrino magnetic moment and neutrino millicharge
  from LUX-ZEPLIN dark matter search results}},
  \href{https://doi.org/10.1103/PhysRevD.107.053001}{\emph{Phys. Rev. D}
  {\bfseries 107} (2023) 053001}
  [\href{https://arxiv.org/abs/2207.05036}{{\ttfamily 2207.05036}}].

\bibitem{Giunti:2023yha}
C.~Giunti and C.A.~Ternes, \emph{{Testing neutrino electromagnetic properties
  at current and future dark matter experiments}},
  \href{https://arxiv.org/abs/2309.17380}{{\ttfamily 2309.17380}}.

\bibitem{XENON:2020rca}
{\scshape XENON} collaboration, \emph{{Excess electronic recoil events in
  XENON1T}}, \href{https://doi.org/10.1103/PhysRevD.102.072004}{\emph{Phys.
  Rev. D} {\bfseries 102} (2020) 072004}
  [\href{https://arxiv.org/abs/2006.09721}{{\ttfamily 2006.09721}}].

\end{thebibliography}\endgroup

\end{document}